\documentclass{article}  
\usepackage{breckenr2005}
\usepackage{graphicx}
\frompage{000} \topage{000}                                              
\def\rightmark{High $p_T$ correlations of $\gamma$ and charged hadrons at RHIC}
\def\leftmark{K. Filimonov}
 
\title{High $p_T$ correlations of $\gamma$ and charged hadrons at RHIC} 
\authors{
{Kirill Filimonov%
}\\[2.812mm]
{\normalsize
Lawrence Berkeley National Laboratory, \\ 
1 Cyclotron Road, Berkeley, CA 94720, USA\\[0.2ex] 
}}
 
\abstract{Prompt photon production in ultra-relativistic heavy-ion collisions 
provides a calibrated probe for the study of the 
properties of high energy density QCD matter. 
Especially interesting are the measurements 
of $\gamma$-tagged jets where the hard scattering scale is known 
and can be used to determine the partonic energy loss in the dense matter. 
We discuss the potential of $\gamma$-jet measurements at 
the Relativistic Heavy Ion Collider (RHIC) and argue that the observed 
supression of the away-side correlations for di-jet production 
in central Au+Au collisions 
at \mbox{$\sqrt{s_{NN}}$}=200 GeV should significantly reduce 
the backgrounds for the $\gamma$-jet coincidence measurements.}

\keyword{} 
\PACS{25.75.-q}
 
\begin{document}
\def\PRL{Phys. Rev. Lett. }
\def\PRC{Phys. Rev. C }
\def\PRD{Phys. Rev. D }
\def\PLB{Phys. Lett. B }
\def\NPA{Nucl. Phys. A }
\def\pT{\mbox{$p_T$}}
\def\pt{\mbox{$p_T$}}
\def\v2{\mbox{$v_2$}}
\def\sqrtsNN{\mbox{$\sqrt{s_{NN}}$}}
\def\sqrts{\mbox{$\sqrt{s}$}}
\def\jt{\mbox{$j_T$}}
\def\kt{\mbox{$k_T$}}
\def\et{\mbox{$E_T$}}
\def\mkt{\mbox{$\langle k_T\rangle$}}
\def\met{\mbox{$\langle E_T\rangle$}}
\def\mjt{\mbox{$\langle j_T\rangle$}}
\def\pizero{$\pi^{0}$}
\def\g{$\gamma$}
\def\etal{{\it et al.}}
\maketitle
\setcounter{page}{1}

\section{Introduction}

In hadronic collisions, direct photons 
(those that do not come from hadronic decays, such as
\pizero$\rightarrow$\g+\g~and $\eta\rightarrow$\g+\g) 
result from parton-parton scatterings with large momentum transfer. They are
predominantly produced from quark-antiquark annihilation ($q+\bar{q}\rightarrow g+\gamma$) and quark-gluon Compton scattering ($q+g\rightarrow q+\gamma$).
Additional non-hadronic-decay component comes from the 
processes in which a photon
is radiated off a quark: $q+X\rightarrow q+X+\gamma$ (bremsstrahlung photons).
The measured direct photon cross section in high-energy hadronic collisions 
is well reproduced in the leading
and next-to-leading order perturbative QCD calculations \cite{Owens:1986mp,Huston:1995vb}.
In ultra-relativistic heavy-ion collisions, thermal direct photons
are expected to be produced in the Quark-Gluon Plasma (QGP) phase 
\cite{Shuryak:1978ij}.
Rates of thermal photons are predicted to be sizeable up to transverse momenta 
\pT$\sim$3-4 GeV/c \cite{Turbide:2003si} 
and their measurements are highly desirable to infer the
temperature achieved in the nucleus-nucleus collisions. 
It was also suggested that partons produced in hard scatterings may interact
with a thermalized parton from the QGP, resulting in a hard+thermal 
component in photon production, which may extend to higher transverse
momenta~\cite{Fries:2002kt}.

Non-thermal photons from hard scatterings 
in heavy-ion collisions are also
 of high interest as they are produced very early in the
collision and escape the hottest and densest stage without further 
interaction. 
The measurements of single-particle inclusive spectrum of 
prompt photons at high \pT~can test the expected scaling of particle 
yields in hard processes with the number of inelastic nucleon-nucleon 
collisions \cite{Reygers:2005sm}.

Energetic partons propagating through the medium are predicted
to lose energy via induced gluon radiation, with the energy loss
depending strongly on the color charge density of the created system and the
traversed path length \cite{quenching}.
Strong suppression relative to $p+p$ collisions of  
inclusive hadron yields at high \pT~measured in central Au+Au collisions
\cite{suppression1,suppression2,Adler:2003qi,suppression3}
is believed to arise from final-state interactions in the dense medium 
that is created at RHIC \cite{dAuPhobos,dAuPhenix,dAuStar,dAuBrahms}. 
The observed suppression is well
described by the pQCD calculations invoking partonic energy loss 
(``jet quenching''). 
However, since the yield of produced hadrons at a fixed \pT~results
from the convolution of the parton production cross section and fragmentation function, suppression of single-particle spectra is not an ideal observable
to measure directly the modification of the fragmentation function 
for a given energy scale. A much better tool to study jet energy loss is
provided by the measurements of \g-jet coincidences, where the inclusive jet
fragmentation function can be extracted from the differential \pT-spectrum of
charged particles in the direction opposite to a tagged direct photon \cite{Wang:1996yh}.

The STAR detector at RHIC \cite{Ackermann:2002ad}
has demonstrated excellent charged particle tracking capability provided by the large acceptance ($|\eta|<$1.3, 2$\pi$ in azimuth) Time Projection Chamber \cite{Anderson:2003ur}. Neutral particle (\g,\pizero) detection in STAR can be performed in full azimuth and $|\eta|<$1, 1$<\eta<$2 ranges covered by the Barrel \cite{Beddo:2002zx} and Endcap \cite{Allgower:2002zy}
segmented electromagnetic calorimeters. 
We have employed the PYTHIA event generator (v. 6.131) \cite{pythia} 
run with default parameters for $p+p$ collisions 
at \sqrts =200 GeV to study the feasibility of measuring \g-tagged jets
using the STAR detector. 
 
\section{Spectra of direct photons}

\begin{figure}[htb]
\vspace*{-.3cm}
                 \insertplot{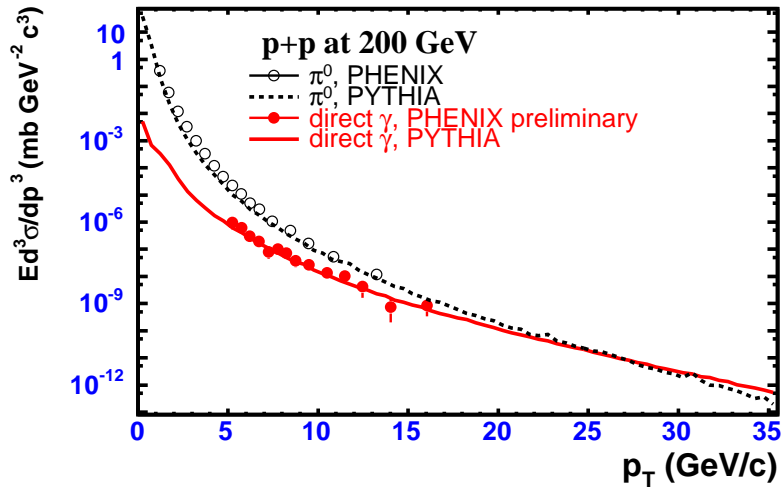}
\vspace*{-1cm}
\caption[]{Measured cross section and PYTHIA calculations for \pizero (open symbols from \cite{Adler:2003pb}, dashed line) and direct-photon (solid symbols from \cite{Okada:2005in}, solid line) production in $p+p$ collisions at \sqrts=200 GeV.}
\label{fig1}
\vspace*{-.3cm}
\end{figure}

First we would like to establish how well the PYTHIA model describes the
experimental data on prompt photon production at RHIC energies.
Measurements of cross section for \pizero~\cite{Adler:2003pb}
and direct \g~production \cite{Adler:2005qk,Okada:2005in}
in  
$p+p$ collisions at \sqrts=200 GeV were reported by the 
PHENIX collaboration.
Figure~\ref{fig1} shows the measured cross sections 
compared to our PYTHIA calculations. 
The model reproduces the data quite adequately, somewhat 
underestimating the measured \pizero~yields at low \pT, 
but shows a better agreement
at higher transverse momenta. In the model, 
\pizero~cross section dominates
over prompt photon production up to \pT$\sim$25 GeV/c. 
So-called isolation cuts which set a limit on the total energy or 
number of particles in a cone around the photon are usually used to reduce the
decay photon background for clean \g-jet identification 
in $p+p$ collisions.

\begin{figure}[htb]
\vspace*{-.3cm}
                 \insertplot{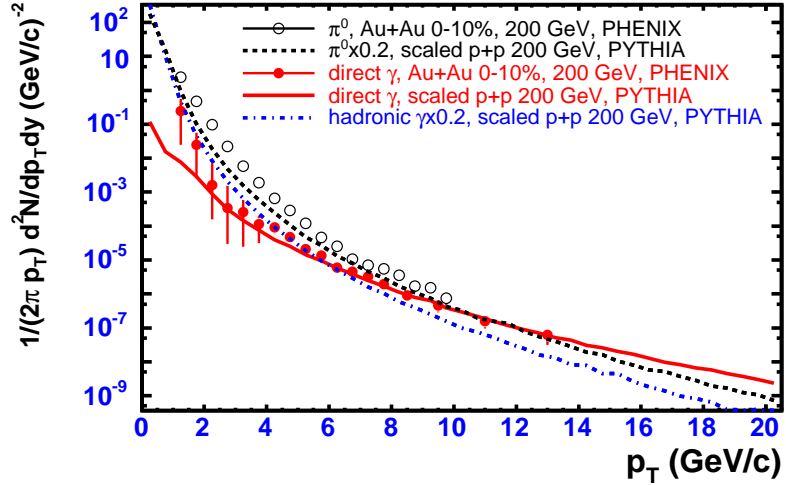}
\vspace*{-1cm}
\caption[]{Measured inclusive spectra and scaled PYTHIA calculations for \pizero (open symbols from \cite{Adler:2003qi}, dashed line) and direct-photon (solid symbols from \cite{Adler:2005ig}, solid line) production in central Au+Au collisions at \sqrts=200 GeV.}
\label{fig2}
\vspace*{-.3cm}
\end{figure}

Due to the high multiplicity of produced particles 
it may be difficult to apply isolation cuts in central 
Au+Au collisions. On the other hand, the \pizero-spectra are suppressed
by a factor of $\sim$5 relative to the scaled $p+p$ collisions, whereas
prompt photon production is shown to be unaffected by the medium \cite{Adler:2005ig}.
Figure~\ref{fig2} compares the inclusive spectra of \pizero~and direct \g~ 
measured in central Au+Au collisions at \sqrts=200 GeV with scaled 
PYTHIA calculations. The \pizero~inclusive spectra from PYTHIA were scaled
by the number of binary collisions and an additional factor of 0.2 to take into
account the experimentally measured suppression in hadron production in
central Au+Au collisions at high transverse momenta. Extrapolating 
experimental data to higher \pT~using PYTHIA, we can see that the yield of 
direct photons is comparable to \pizero~yield around
\pT$\sim$12 GeV/c. Moreover, the direct photon production dominates
yields of photons from hadronic decays above \pT$\sim$8-10 GeV/c.

\section{Jets tagged by direct photons}
Hard scattered partons fragment
into a high energy cluster (jet) of hadrons which are 
distributed in a cone of size $\Delta\eta\sim\Delta\phi\sim0.7$ 
in pseudorapidity and azimuth.
Correlations of high \pT~hadrons
were successfully used for the identification of jets on a statistical basis
in Au+Au and $p+p$ collisions at RHIC ~\cite{flow,btob}.
The relative azimuthal angle distributions of di-hadrons reveal jet-like 
correlation characterized by the peaks at $\Delta\phi=0$ 
(near-side correlations) 
and at $\Delta\phi=\pi$ (back-to-back).  
Direct photons produced in hard scatterings are accompanied by a jet of 
hadrons in the azimuthal direction opposite to that of the photon.
High \pt~\g-hadron relative azimuthal angle distributions 
are then expected to produce only away-side correlations.

\begin{figure}[htb]
\vspace*{-.3cm}
\begin{flushleft}{\includegraphics*[%
  height=1.8in,
  width=0.49\columnwidth]{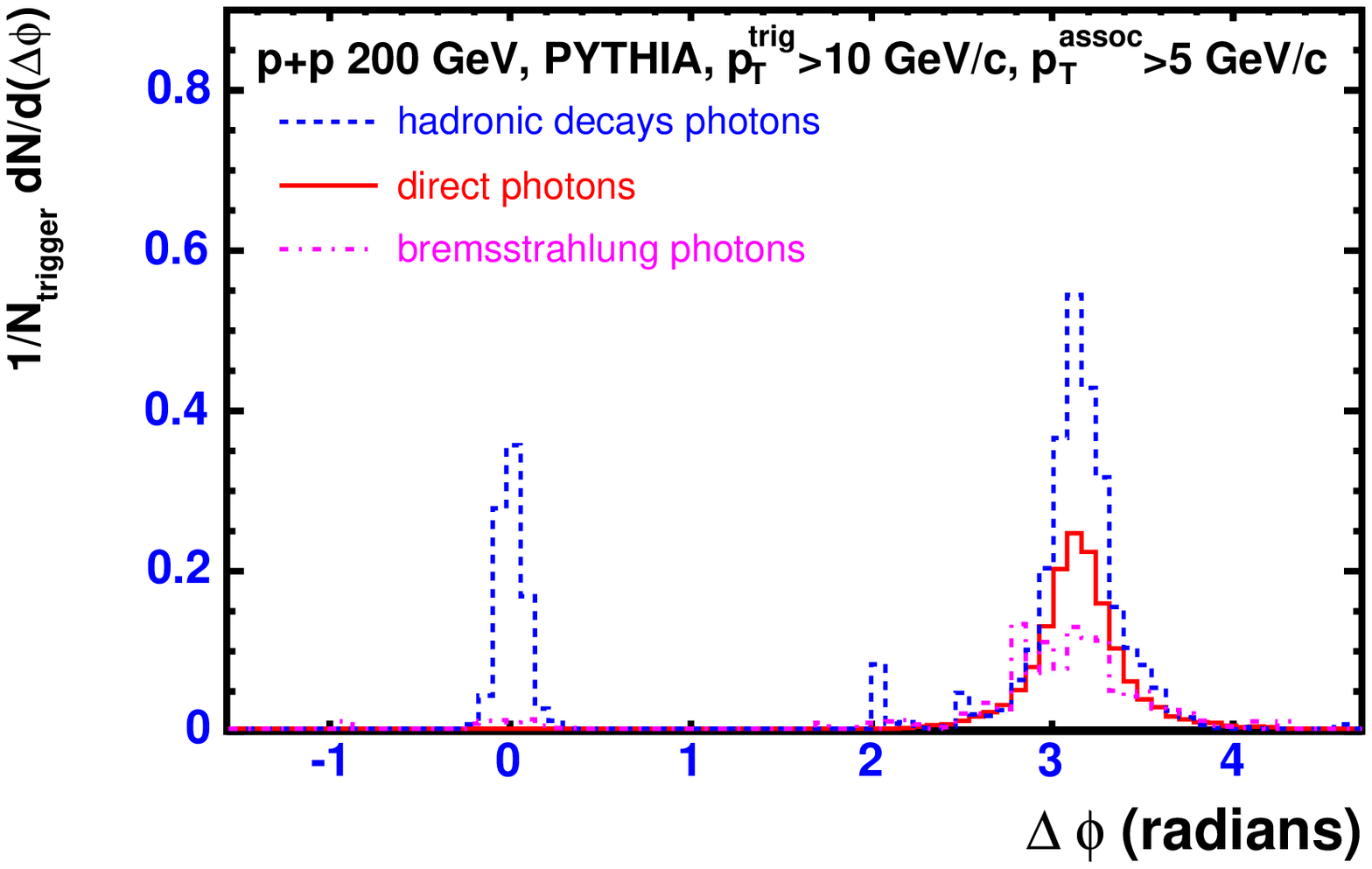}}
{\includegraphics*[%
  height=1.8in,
  width=0.49\columnwidth]{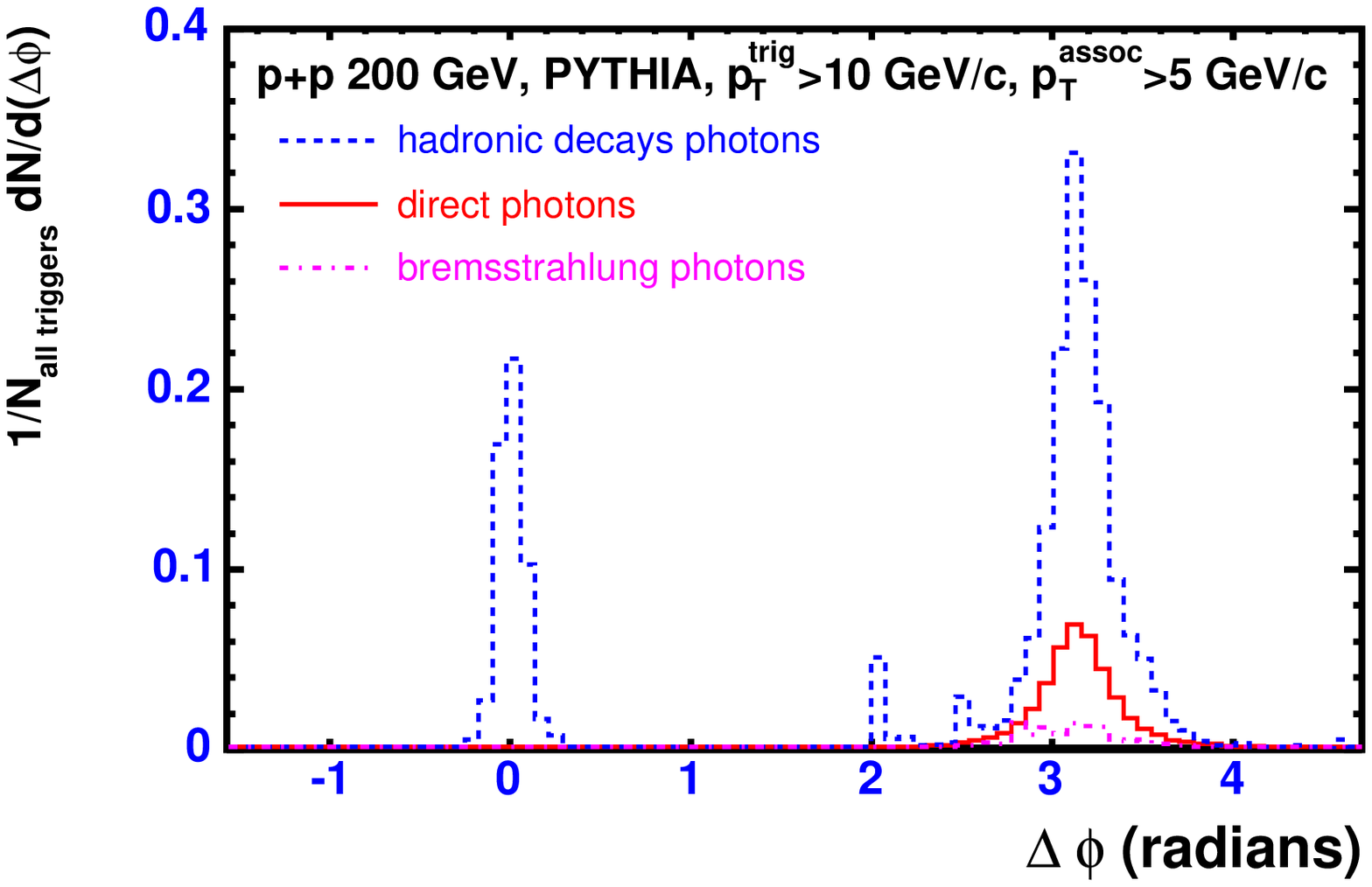}}\end{flushleft}
\vspace*{-1cm}
\caption[]{PYTHIA calculations of the relative azimuthal angle distributions between photons with \pt$>$10 GeV/c and charged hadrons with 5$<$\pt$<$10 GeV/c 
in $p+p$ collisions at \sqrts=200 GeV. Left panel: distributions are normalized separately for hadronic decay, direct, and bremsstrahlung photon triggers. Right panel: distributions are normalized to the total number of photon triggers.} 
\label{fig3}
\vspace*{-.3cm}
\end{figure}

Figure~\ref{fig3} shows the PYTHIA calculations of the relative azimuthal angle distributions between photons with \pt$>$10 GeV/c and charged hadrons with 5$<$\pt$<$10 GeV/c 
in $p+p$ collisions at \sqrts=200 GeV (photons and charged hadrons were taken
within $|\eta|<$1 corresponding to the STAR experimental acceptance).
Left panel of Figure~\ref{fig3} shows the distributions normalized separately for hadronic decay, direct, and bremsstrahlung photon triggers.
Photons coming from hadronic decays (mostly \pizero's) are correlated
to the charged hadrons on the near and away sides. Their correlation is 
stronger (larger associated hadron multiplicity) than that of direct photons,
since hadronic decay photons originate from higher \pT~neutral pions, which
in turn come from fragmentation of yet higher $E_T$ partons. 
In the \g-jet measurements, bremsstrahlung photons
are background to the study of the processes where energy of the photon
precisely characterizes the momentum transfer $Q^2$ of the hard scattering.
In PYTHIA, photons coming from the processes involving initial/final state
radiation contribute 40-30\% to the
total yield of prompt photons in the range \pt=6-12 GeV/c, 
in agreement with the 
NLO pQCD calculations \cite{Gordon:1993qc}.
Their correlation with the high \pT~charged hadrons 
is predicted to be weaker than that of the true direct photons.
Right panel of Figure~\ref{fig3} shows the distributions normalized to the
total number of photon triggers from all sources. The direct \g-jet 
correlations are predicted to contribute 
less than 20\% to the total \g-charged distribution in $p+p$ collisions 
with the selected $p_T^{\rm trig}>$10 GeV/c and $p_T^{\rm assoc}=$5-10 GeV/c.

In central Au+Au collisions, dihadron correlations are strongly affected
by in-medium effects.
Correlations of high \pT~hadrons at small relative angles are seen to be 
essentially 
unaffected by the medium (the strength of the near-side correlations 
is consistent with that measured in p+p and d+Au collisions).
In sharp contrast, the away-side (back-to-back) correlations are 
strongly suppressed in the most central Au+Au collisions \cite{dAuStar,btob}.
Back-to-back suppression also varies with azimuthal orientation of 
the jets relative to the reaction plane for non-central Au+Au collisions,
exhibiting larger
suppression of the back-to-back correlations for the 
out-of-plane trigger particles
than for in-plane \cite{inoutStar}.
These observations are naturally predicted by jet quenching 
models, where the energy  
loss of a parton depends on the density of and distance traveled 
through the medium.
The high \pt~trigger biases the initial production point to be near 
the surface so the near-side correlations should be similar to those
seen in p+p collisions. The away-side correlations are suppressed 
in the dense medium, and more suppressed 
when the trigger hadron is emitted perpendicular to the reaction plane.

\begin{figure}[htb]
\vspace*{-.3cm}
                 \insertplot{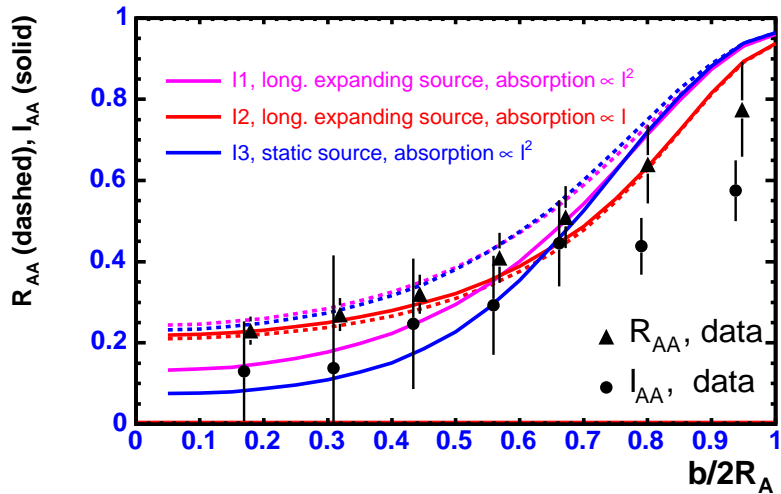}
\vspace*{-1cm}
\caption[]{Measurements of the nuclear modification factor $R_{AA}$ (triangles,
\pizero's with \pt$>$4 GeV/c, from \cite{Adler:2003qi}) 
 and away-side suppression 
factor $I_{AA}$ (circles, charged hadrons with $p_T^{\rm trig}=$4-6 GeV/c and $2<p_T^{\rm assoc}<p_T^{\rm trig}$ GeV/c, from \cite{btob}) in Au+Au collisions at \sqrtsNN=200 GeV compared
to different jet absorption model scenarios as a function of the impact parameter $b$
normalized by the nuclear diameter $2R_A$.} 
\label{fig4}
\vspace*{-.3cm}
\end{figure}

\begin{figure}[htb]
\vspace*{-.3cm}
                 \insertplot{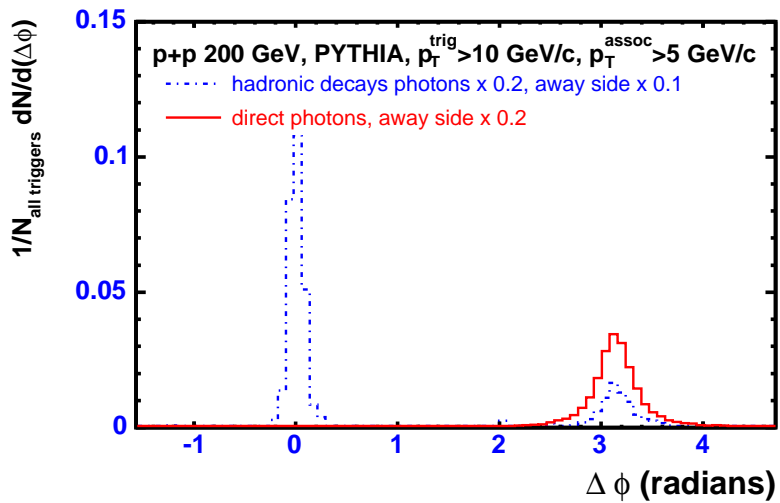}
\vspace*{-1cm}
\caption[]{Scaled distributions from the right panel of Figure~\ref{fig3}. The distributions are normalized to the total number of hadronic+direct photon triggers.} 
\label{fig5}
\vspace*{-.3cm}
\end{figure}

For \g-jet processes in heavy-ion collisions, there is no bias in
the initial production point as direct photons are expected to 
escape the medium without interaction. The accompanying back-to-back
parton will travel through the medium and, in jet quenching models,
will lose energy, resulting in the suppression of the away side correlations
with respect to the reference nucleon-nucleon system.
This suppression, however, should be smaller than that in the case of 
di-jet production, where the away side parton propagates through the
largest distance through the dense medium.
In \cite{btob}, the suppression of the away-side di-hadron correlations
has been quantified using $I_{AA}$ parameter. Similarly to the
nuclear modification factor $R_{AA}$, taken as the ratio of single-particle
yields measured in $AA$ collisions to the geometry-scaled yields measured in
nucleon-nucleon collisions, $I_{AA}$ is defined as the ratio of the 
di-hadron correlation strength measured in $AA$ to the reference $p+p$ data.
$I_{AA}$ for the away-side correlations was found to be smaller or comparable 
to $R_{AA}$, indicating that the away-side jets might be suppressed more 
strongly 
than single-inclusive \pT-distributions, in agreement with the premise
of the surface emission bias. 
Recently, a simple medium induced jet absorption model 
incorporating realistic nuclear geometry was used to quantitatively 
describe the centrality
dependence of the observed suppression of the high \pt~hadron yield and of the back-to-back angular
correlations \cite{Drees:2003zh}. We performed very similar calculations
shown in Figure~\ref{fig4}.
In the jet absorption model, the
geometry of the two overlapping nuclei with the 
Woods-Saxon density distribution is calculated 
using a Monte Carlo Glauber approach.
Jet energy loss is modeled using an absorption coefficient and 
the matter integral along the path of the parton, parameterized 
for the different dependencies of the absorption on the path length.
Parameterizations assuming
the $l^2$-dependence of the absorption on the path length $l$ describe
the experimental data quite well (Figure~\ref{fig4}).
A generic prediction of the model is that the back-to-back correlations
are more suppressed than the single-particle production, by as much as a
factor of 2. If this holds true, the \g-jet measurements at RHIC should
be easier to perform. Since there is no initial production ``surface''
or ``skin'' bias associated with selecting \g-triggers, the \g-accompanying
jet will be suppressed by a single-particle rate, which is less than the
di-hadron suppression rate for jet-jet processes.
It is illustrated in Figure~\ref{fig5}, where we show the
distributions from the right panel of Figure~\ref{fig3}, but with hadronic 
\g~production scaled by the \pizero-suppression factor
of 0.2, and assuming a back-to-back suppression factor
of 0.1 for the away side. For the direct \g~distribution, the back-to-back suppression is assumed to be equal to the  \pizero-suppression factor.
For the selected \pT-ranges of the trigger
photon $p_T^{\rm trig}>$10 GeV/c and associated charged hadron $p_T^{\rm assoc}=$5-10 GeV/c, the combinatorial background even in the most central Au+Au 
collisions is expected to be negligible, considering the product of the
single-particle per-event yields at those transverse momenta.
On the other hand, the partonic energy loss is predicted to only weakly depend
on the energy of the parton.
Assuming that the di-hadron back-to-back suppression is twice the
single-particle suppression and using PYTHIA to simulate the
jet fragmentation at \sqrts=200 GeV, we estimate that up to 70\% of the
\g-charged back-to-back azimuthal correlations in central Au+Au collisions
may come from direct \g-jet events. The situation can further be improved 
using higher \pT~for the trigger photon and associated hadrons. Considering the integrated luminosity achieved
in the last RHIC run of Au+Au collisions at \sqrtsNN=200 GeV, and
using the PHENIX measured and PYTHIA-extrapolated cross section
for direct photon production, we conclude that the exploratory \g-jet
measurement may already be achievable with the STAR data on tape.

\section*{Acknowledgment(s)}
This work was supported by the Director, Office of Science, Nuclear Physics, 
U.S. Department of Energy under Contract DE-AC03-76SF00098.


\vfill\eject
\end{document}